\documentclass{aa}  
\usepackage{graphicx}
\usepackage{ulem}
\usepackage{txfonts}
\usepackage{xcolor}

\begin{document}

\title{Magnetic precession as a model for QPOs in PULXs:} 
\subtitle{flat-top noise ULXs are different}

   \author{Miljenko \v Cemelji\'c\inst{1,2,3,4}
          \and
          W{\l}odek Klu\'zniak
          \inst{1,2}
          \and 
          Sukalpa Kundu
          \inst{2}
          }
 \institute{Nicolaus Copernicus Astronomical Center of the Polish Academy of Sciences, Bartycka 18, 00-716
Warsaw, Poland\\
\email{miki@camk.edu.pl}
\and
Nicolaus Copernicus Superior School, College of Astronomy and Natural Sciences, Gregorkiewicza 3, 87-100, Toru\'{n}, Poland
\and
 Research Centre for Computational Physics and Data Processing, Institute of Physics, Silesian University in Opava, Bezru\v{c}ovo n\'am.~13, CZ-746\,01 Opava, Czech Republic
\and
Academia Sinica, Institute of Astronomy and Astrophysics, P.O. Box 23-141,
Taipei 106, Taiwan
}

   \date{\today}

 
  \abstract
   {Several instances of low frequency  Quasi-Periodic Oscillations (QPOs) have been reported in ultraluminous X-ray sources (ULXs), including three in pulsating ones (PULXs) to date. The nature of many ULXs is still unclear, as are the detailed properties of accretion in PULXs. }
   {We seek an answer to questions such as: Is there a QPO model that fits the data? Can mHz QPOs  be used to constrain the magnetic field and accretion rate of the neutron stars in PULXs? Are all the low frequency QPOs in ULXs a manifestation of the same phenomenon?}
   {We apply Dong Lai's precession model to the PULX data, with the magnetic threading of the accretion disk constrained by recent simulations.}
   {Based on the magnetic precession model, and on recent progress in understanding the inner structure of accretion disks, we predict an inverse scaling of QPO frequency with the neutron star period in PULXs. The theoretical curve is largely independent of the stellar magnetic field or mass accretion rate and agrees with the data for the known QPOs in PULXs. The flat-top QPOs detected in ULXs have observational properties that seem to be very different from the QPOs detected in PULXs, indicating they might have a different origin.}
   {}

   \keywords{Stars: neutron -- X-rays: binaries -- Accretion, accretion disks
               }

   \maketitle
%

\section{Introduction}

Ultraluminous X-ray sources (ULXs) are extragalactic off-nuclear, point-like sources with extremely high X-ray luminosities, typically in the range $L_{\rm X} \approx 10^{39}-10^{41} \rm \ erg\ s^{-1}$ \citep{Fabbiano+89,RP1999,RW2000}. The observed emission from these sources generally exceeds the Eddington limit of stellar-mass objects \citep{Makishima+00,King+01}. Recently, \citet{Tranin+24} reported 1901 such objects using Chandra, XMM-Newton, and Swift data. As a result of the super-Eddington luminosities, they were initially classified as intermediate-mass black holes (IMBH) \citep{Kaaret+06,Kong+07}. However, stellar evolution theories generally do not predict the formation of such objects, and the observed abundance is difficult to produce even in dense star clusters \citep{King+04,Madhusudhan+06}. A significant breakthrough in discovering the true nature of ULXs occurred when \citet{Bachetti+14} discovered a coherent pulsation with an average period of $\approx1.37$ s from the source ULX M82 X-2. This strongly suggested that the compact accretor was a neutron star. Since then, several ULXs have been observed with a coherent pulsation \citep[see e.g.,][]{Furst+16, Israel+Sc17,Carpano+18,Castillo+20,Pintore+25}. The IMBH scenario has been largely ruled out for these sources.

The challenge to explain the extreme super-Eddington luminosities for stellar-mass compact objects was met by \citet{King+01} who proposed an explanation in which beamed emission could boost the observed luminosity of stellar mass objects to very high values. As an extension of this idea, \citet{King+17} proposed a mechanism---the KLK model---that accounts for the extremely high apparent luminosity of these pulsars within the standard Shakura-Sunyaev framework \citep{SS73}. In the KLK model an analysis of spin-up torques yields moderate magnetic fields of $B\sim10^{11}\,$G. Alternatively, a strong magnetic field ($B>10^{14}$ G) has been proposed  to reduce the opacity, which would allow the true luminosity to be several times the Eddington limit \citep{P92,Mushtukov+15}. 

Although fairly rarely observed in ULXs, quasi-periodic oscillations (QPOs) from these sources provide us with another distinct timescale alongside pulsations that helps to probe deeper into the accretion process. Several examples of these oscillations have been reported in the ULX sources over the past few years \citep[see e.g.,][]{Feng+10,Hasan+14,Agarwal+15,Urquhart+22,Imbrogno+24}. Among these, only a few sources have reported measurements for both the pulsation period and QPO frequency, currently these are M82 X-2 \citep{Feng+10}, ULX-7 M51 \citep{Imbrogno+24}, NGC 7793 P13 \citep{Imbrogno24talk}.

In this work, we revisit the precession model for tilted accretion disks that was initially proposed by \citet{Lai99} and verified numerically in \citet{Pfeiffer+04}. We apply general scaling laws to restrict the model. Refining the precession frequency equation, we propose an inverse relationship between the QPO frequency and the neutron star rotation period (Section 2). A comparison of this prediction with the known QPOs in PULXs (desribed in Section 3) is carried out in Section 4, and their properties are contrasted with those of flat-top noise QPOs in ULXs.

\section{Dong Lai's magnetic precession model and QPOs}
\label{S2}
We start with the formula for the precession frequency of a ring in a tilted accretion disk \citep{Lai99}, as summarized in \cite{Pfeiffer+04} in their numerical work demonstrating the warping, tilting, and the rigid-like precession of an accretion disk subjected to the magnetic forces exerted by an inclined stellar magnetic dipole. The precession frequency is given by
\begin{equation}
\nu(r)=\frac{\mu^2}{2\pi^3r^7\Omega(r)\Sigma(r)D(r)}F(\theta).
    \label{prec}
\end{equation}
Here, $F(\theta)=2f\cos^2(\theta)-\sin^2(\theta)$ is a function of the (fixed) angle between the stellar spin axis and the axis of the (centered) magnetic dipole of magnitude $\mu$, while $f$ is a factor describing the degree to which the accretion disk is threaded by the magnetic field of the dipole. 

We take $f=1$, corresponding to no threading of the disk, in agreement with the analytic calculations of \cite{Parth23}, confirmed by their numerical simulations, which showed that the magnetic field inside the disk is strongly suppressed with respect to the value above and below the disk. Hence, we take $F(\theta)=2\cos^2(\theta)-\sin^2(\theta)$, which varies between the value of 2 for $\theta=0$, and 5/4 for  $\theta=30^\circ$. Thus, we expect the value of $F$ to vary by a factor of up to about 2 from source to source in beamed ULXs.

We assume the accretion disk is geometrically thick for ULX accretion rates, in agreement with the simulations of \cite{Abarca21, Inoue23, Inoue24, Kayani25}. The quantity $D(r)$ in Eq.~\ref{data} is a geometrical factor that close to the inner edge of the disk then becomes $\sqrt{2H/r}$, with $H$ the half-thickness of the disk. Without further ado, we take $D(r) = \mathrm{const} \approx 1$. Thus, for any given source we take $F(\theta)/D(r)=\mathrm{const}\approx 1$.

We identify the QPO frequency in PULXs  with the frequency $\nu(R)$ of Eq.~\ref{data}, where $R$ is a characteristic radius (discussed below) of the inner precessing ring. The surface density of the disk, $\Sigma$, is unobservable, but it can be replaced with the mass accretion rate $\dot M$, through the mass conservation equation $\dot M\approx -2\pi r \varv_r \Sigma$, where $\varv_r$ is the radial velocity. Our crucial step is to assume that $\varv_r(R)$ scales as the azimuthal velocity, $R\Omega(R)$, in the inner disk, 
\begin{equation}
\varv_r =-K' R\Omega.
\label{kprime}
\end{equation}
The constant $K'$ defined by this equation is approximately equal to the alpha viscosity parameter, $K'\approx \alpha$, on the assumption that $H\sim R$ \citep{KK00}. We will change this approximate equality into an exact one introducing a new parameter $a\approx 1$, so that $K'=a\alpha$.
The QPO frequency becomes
$$
\nu_\mathrm{QPO}=K'\frac{\mu^2}{\pi^2 R^5\dot M}\frac{F}{D},
$$
with $F/D\approx 1$.
As our final assumption, we take the characteristic radius, $R$, to be somewhat larger than the radius of the inner edge of the disk, taken to be magnetically truncated below the corotation radius. Thus, from
\begin{equation}
R/d= R_{\mathrm{in}}=\tilde\eta\left(\frac{GMP^2}{4\pi^2}\right)^{1/3}= \eta\left(\frac{\mu^4}{GM\dot M^2}\right)^{1/7},
\label{arr}
\end{equation}$\mu^2/\dot M=2\pi\, \tilde\eta^{-3/2}\eta^{-7/2}(R/d)^5/P$, where $P$ is the neutron star rotation period, and the parameters satisfy  $d>1$, $\tilde\eta<1$, while the constant $\eta\approx 1$ takes into account the theoretical uncertainty of up to about a factor of 2 \citep{KluRap07} in the value of the magnetic truncation radius (the last term in Eq.~\ref{arr}).
Thus, finally, with our assumptions we obtain a simple inverse proportionality between the QPO frequency and the stellar spin period, $P$,
\begin{equation}
\nu_\mathrm{QPO}=K /P,
    \label{qpo}
\end{equation}
where the proportionality constant
\begin{equation}
    K=(2/\pi)\tilde\eta^{-3/2}\eta^{-7/2}d^{-5}(F/D)K',
    \label{konst}
\end{equation}
with the $K'$ of Eq.~\ref{kprime} given by $K'=a\alpha$, as already noted.
Taking as fiducial values $(2/\pi)\eta^{-7/2}=1$ , $a=1$,   $(F/D)=1$, we obtain $K =\tilde\eta^{-3/2} d^{-5}K'=\tilde\eta^{-3/2}d^{-5}|\varv_r|/(r\Omega)= \alpha\tilde\eta^{-3/2}d^{-5}$.

As $|\varv_r|<<r\Omega$, we necessarily have $\nu_\mathrm{QPO}/P<<1$. In fact, since $\alpha\sim 0.01$ according to magneto-rotational instability (MRI, \cite{BH91a}) simulations,  we would expect the QPO frequency to be about two orders of magnitude smaller than the spin period when corotation ($\tilde\eta =1$) holds,
$\nu_\mathrm{QPO}\sim 0.01/ P$.  
 This seems to be roughly the case (Table 1). Assuming the center of the precessing torus to be at a radius some 30\% larger than the truncation radius $R_{\mathrm{in}}$ we would take $d=1.3$, a value giving QPO frequencies
\begin{equation}
    \nu_\mathrm{QPO}\approx 0.003/ P,
    \label{fit}
\end{equation}
that provide a reasonable fit to the still very limited data, as seen in Fig.~\ref{FigVibStab}.

\section{mHz QPOs in PULXs}
We are aware of only three PULXs for which mHz QPOs have been reported: M51 ULX7, M82 X-2, and NGC7793 P13; their spin periods and approximate QPO frequencies have been listed in  Table~\ref{data}. 

\subsection{M82 X-2}
The X-ray source X42.3+59 was identified as an ultraluminous X-ray source (ULX)  using Chandra X-ray observations \citep{Kaaret+06,Kong+07} and it was subsequently named M82 X-2. \citet{Kaaret+06} found significant timing noise at 1 mHz from this source, subsequently QPOs in the range in the range 2.77--3.98 mHz were reported by \citet{Feng+10}. \citet{Bachetti+14} discovered a pulsation from this source with an average period of $1.37\,$s and a sinusoidal modulation of around 2.5 days; the nearly-circular orbit with eccentricity $<0.003$ and the high luminosity ($1.8 \times 10^{40}\,\mathrm{erg}\,\mathrm{s}^{-1}$) suggested accretion through Roche lobe overflow. During the interval MJD 56696 to 56701 the spin-up rate was $\dot{P} \approx -2 \times 10^{-10} \mathrm{ss}^{-1}$. \citet{Bachetti+22} found spin-down during the period 2016-2020 and stated that the spin history suggests that the source is near spin equilibrium. \citet{Liu+24} followed the evolution from the last 20 years to find a gradual spin-down, with occasional spin-up events in between. \citet{Bachetti+22} suggest a $B > 10^{13}\,$G magnetic field to explain the high luminosity of about 200 times the Eddington limit. However, from the spin torques, $B\approx10^{11}\,$G and a beaming factor of $b\approx0.06$ is found in the KLK model \citep{King+17}. 


\subsection{M51 ULX-7}
First discovered as an X-ray source by \citet{RW2000} in the ROSAT survey, it was observed to have a $2.1\,$h periodicity \citep{Liu+02} in a Chandra ACIS observation. \citet{Earnshaw+16} found a strong variability ($>10\%$) in this source, even in the hard state (photon index $\Gamma\approx1.5$), which is uncommon in ULXs. The pulsar nature of the compact object was demonstrated when \citet{Castillo+20} detected a pulsation at $\approx2.8$ s using data from the XMM-Newton Large Program UNSEeN. They suggested this source to be a High Mass X-ray Binary (HMXB) with luminosity varying from $L_{\rm X} < 3\times 10^{38} \mathrm{erg}\,\mathrm{s^{-1}}$ to $L_{\rm X} \approx 10^{40} \mathrm{erg}\,\mathrm{s^{-1}}$. They also measured a spin-up rate of $\dot{P}\approx - 1.5\times 10^{-10} \mathrm{s}\,\mathrm{s^{-1}}$ during the two months 2018 May and June. Later, \citet{Brightman+22} followed up the spin evolution in the period 2018-19 to report an average spin-up rate of $\dot{P}\approx - 3\times 10^{-10} \mathrm{s}\,\mathrm{s^{-1}}$. However, a long-term (2005-2018)  secular spin-up rate of $\dot{P}_{\rm sec} \approx -10^{-9} \mathrm{s}\,\mathrm{s^{-1}}$ was also reported by \citet{Castillo+20} in the same study. Recently, quasi-periodic oscillations in the frequency range $0.449-0.565$ mHz were discovered by \citet{Imbrogno+24}. This made ULX-7 the second object from which QPOs have been detected from a PULX accreting at a super-Eddington luminosity. While \citet{Vasilopoulos+19} estimated the surface magnetic field to be close to $2-7 \times 10^{13}\,$G based on accretion torques, \citet{Castillo+20} inferred a wider range of $B\approx8\times10^{11}-10^{13}\,$G considering both the effects of spin-up torques and column accretion \citep{Mushtukov+15}; in the same study, the beaming factor was estimated to lie between $1/12<b<1/4$. However, applying the KLK model, \cite{KingLasota20} obtain $b=0.09$ and $B$
between $6.9\times10^{9}\,$G and $1.9\times10^{11}\,$G, depending on the spin-up rate.

\subsection{NGC 7793 P13}
This object was first observed as a highly luminous ($L_X\approx10^{39}\mathrm{erg}\,\mathrm{s^{-1}}$) variable X-ray source by \citet{RP1999} in a ROSAT survey. \citet{Motch14} detected this to be a binary system with a compact accretor and a B9Ia donor star, with an orbital period of $\approx64$ days. \citet{Furst+16} discovered a nearly sinusoidal pulsation at $\approx0.42$ s, confirming the accretor to be a neutron star. The luminosity they obtained was close to $50L_{\mathrm{Edd}}$, however, the beaming factor necessary to explain this observation, $b\approx1/50$ was thought to be incompatible with the smooth sinusoidal pulsations. During the period 2013-2016, they detected a spin-up at a rate of $\dot{P}\approx-3.486\times 10^{-11} \mathrm{s}\,\mathrm{s^{-1}}$ from this source.  \citet{Furst21} followed the spin evolution to find a continuous spin-up up to 2020, after which it went into a low state. A similar phase also occurred previously in the period 2011-2013. \citet{Furst+24} followed up the source to discover that it reemerged from the low state in 2022, with the spin-up rate accelerated to $\dot{P}=-6.28\times 10^{-11} \mathrm{s}\,\mathrm{s^{-1}}$.  A QPO at $\approx10\,$mHz was recently reported from this object by \citet{Imbrogno24talk}. \citet{Furst+16} estimated the surface magnetic field of this source to be around $B\approx 5\times10^{12}$ G, based on the standard disk accretion model from \citet{GL79}. However, in the framework of the KLK model \citep{King+17}, 
a surface magnetic field of $B\approx10^{10}$ G could be sufficient to describe the observed luminosity, with a beaming factor $b\approx0.18$. This is the only known PULX with an optical counterpart to date \citep{Motch14}. It has the shortest period among the known PULX sources \citep{Israel+17}.

   \begin{table}
      \caption[]{PULX mHz QPOs}
     $$ 
         \begin{array}{p{0.3\linewidth}p{0.2\linewidth}p{0.25\linewidth}}
            \hline
            \noalign{\smallskip}
            Source      & spin Period & QPO frequency  \\
            \noalign{\smallskip}
            \hline
            \noalign{\smallskip}
            M51 ULX7 & 2.8 s   & 0.5--0.6 mHz   \\
           M82 X-2 & 1.3 s & 2.8--4.0 mHz      \\
            NGC7793 P13 &  0.4 s  & ~~$\approx 10$ ~~mHz \\
            \noalign{\smallskip}
            \hline
         \end{array}
     $$ 
         \label{data}
   \end{table}
   \begin{figure}
   \centering
   \includegraphics[width=\columnwidth]{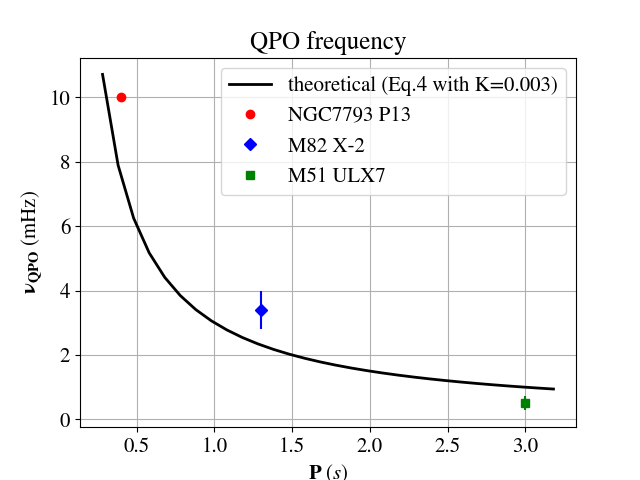}
      \caption{A $\nu_\mathrm{QPO}=\mathrm{const}\cdot 1/P$ line (in solid black) is overlaid on the observational data points with error bars in the cases of M82 X-2 and M51 ULX7, and a preliminary value in the case of NGC7793 P13.
               }
    \label{FigVibStab}
   \end{figure}

\section{Results and discussion}

In Fig.~\ref{FigVibStab} we show the data points for the three PULX QPOs. We also show the theoretical curve of Eq.~\ref{qpo}. Since the actual value of the $\alpha$ parameter is uncertain, we fix the K constant at a reasonable value of $K=0.003$. An interpretation of this value of $K$ is offered in the discussion of Eq.~\ref{fit}. Note that the data are within a factor of 2 of the curve. 

\subsection{The QPO model for PULXs}

We apply Dong Lai's magnetic precession model \citep{Lai99} to ULXs, identifying the precession frequency with that of the observed QPOs.  \cite{Pfeiffer+04} have shown numerically that the inclined dipole will warp the disk and force it to move in a steady, rigid-body like precession of the inner region, with frequency given by Eq.~\ref{prec}.
The model assumes the accreting star to be endowed with a dipole moment of strength $\mu$ that is inclined (at an angle $\theta$) to the rotation axis of the star. The disk is assumed to be truncated by the magnetosphere at radius $R_{\mathrm{in}}$, so necessarily we assume that the radius of the star is smaller: $R_*<R_{\mathrm{in}}$.

As it turns out, neither the radius, $R_*$, nor the mass, $M$, of the neutron star enters the final formula for the QPO frequency. However their values matter overall: they control the luminosity in the inner region,  given by $L=GM\dot M/R_*$ for a given accretion rate $\dot M$ through the inner regions of the accretion disk of an essentially non-rotating star, thus indirectly shaping  the unperturbed shape of the inner disk. Assuming that the accretion disk is fairly thick ($H\sim r$) we were able to transform Eq.~\ref{prec} to the simple form of Eq.~\ref{qpo}. In doing so, we introduced a number of parameters, each of value close to unity. They describe the unknown details of the system geometry, the physics of disk truncation, and the formation of the precessing torus. Together with the physical parameters describing the system ($\mu$, $\theta$, $P$, $\dot M$), and the accretion physics ($\alpha$), they come in three categories.

   \begin{enumerate}
      \item Parameters that are fixed once and for all.
      \item Parameters that have fixed values for a given source.
      \item Parameters that may vary with time.
   \end{enumerate}

The first category includes $\eta$ of Eq.~(\ref{arr}). It determines the exact value of the truncation radius as a function of the physical parameters of the system, and is assumed to be the same for all disks to which we apply the last formula of Eq.~(\ref{arr}). Its value in the literature varies from about 0.4 to about 1 \citep[as reviewed in][]{KluRap07}. The other parameter in this category  is $a$ which, multiplied by the $\alpha$ viscosity, defines the ratio of radial to azimuthal velocities in the disk $\varv_r/(r\Omega)$. The proportionality ``constant'' $K$ in the final formula, Eq.~\ref{qpo}, depends linearly on $a\alpha$. We use the value $a\alpha=0.01$ and take it to be fixed for all the discussed sources, assuming the disk to be in the same state (it is conceivable that $a$ or $\alpha$ may change in a disk state transition). Finally, while the factor $D$ depends on the exact shape of the inner disk, we take it to be the same for all sources, again assuming their accretion disks to be in the same state.

The second category is comprised of the two parameters defining the strength of the stellar magnetic dipole, $\mu$, and its inclination $\theta$. Their influence on the constant $K$ is very weak.
As discussed in Section \ref{S2}, $\theta$ may affect the QPO frequency by a factor of about 2, from source to source. The weak dependence of $K$ on $\mu$ is discussed below.

The most important parameter changing from source to source and (possibly) with the accretion rate is the factor $d$ in Eq.~(\ref{arr}). It is a measure of the radial extent of the precessing warped region of the inner disk. It must be related to the strength of the QPO in the light curve, and it cannot be too close to unity if the QPO is to be observable with the current instruments. For the purposes of this paper we take $d=1.3$, but it could as well be 1.1, or 1.5. At present, there is no theory of what would determine the size of the precessing region. The constant $K$ in Eq.~(\ref{arr}) depends rather sensitively on the value of $d$, being inversely proportional to its fifth power. Given this, it is remarkable that a single curve fits the data so closely (Fig.~\ref{FigVibStab}). 

The last category includes the physical parameter $\dot M$ and the $\tilde\eta$ parameter in our formulae. The former enters the latter through the $\mu^2/\dot M$ combination, with $\tilde\eta\propto (\mu^2/\dot M)^{2/7}$.
Note the weak dependence (inverse two seventh power) on the mass accretion rate. The $\tilde\eta$ parameter defines how far from spin equilibrium the source is (at any given moment), so in principle it can be determined from the spin history with no knowledge of the magnetic moment or of the mass accretion rate. In this sense, the observed QPO frequencies do not (further) constrain $\mu$ or $\dot M$ in this model, although the $\tilde\eta\propto\mu^{4/7}\dot M^{-2/7}$ relation may provide a consistency check of the model. 
The further away from spin equilibrium is the neutron star, the further away from corotation with the star the inner edge of the disk must be, and hence,  by Eqs.~(\ref{qpo}) and (\ref{konst}), the larger the QPO frequency. For a given neutron star, the model gives $\nu_\mathrm{QPO}\propto \dot M^{3/7}$ for a fixed value of the $d$ parameter.
\subsection{Flat-top ULXs}

We would like now to turn to another class of QPOs. These are the sub-Hz QPOs in ULXs showing a flat-top low frequency noise and no coherent pulsations. \cite{Atapin19} report on the properties of flat-topped noise and 0.01 Hz to 1 Hz QPOs in several ULXs, none of which is known to exhibit coherent pulsations. A striking feature of those QPOs is that their frequency in individual sources typically varies by a factor of up to 4 or more, often in strong correlation with the count rate. In three of the six sources reported the QPO frequency  exhibits a very strong dependence on the count rate in {\sl individual sources}, following  a power law dependence with index between 3 and 4 \citep[c.f. figure 4 of][]{Atapin19}. This strongly correlated variability is not as yet explained in any model of QPOs.

The QPO frequency increasing as roughly the third power of the count rate  would be difficult, but perhaps not impossible, to understand in terms of the magnetic precession model.  As remarked earlier, in slowly rotating neutron stars the luminosity (and presumably the reported count-rate) is given by $GM\dot M/R_*$, so it depends on $\dot M$ alone.
However, as we have just seen in the previous subsection, the magnetic precession model applied to PULXs yields $\nu_\mathrm{QPO}\propto \dot M^{3/7}$. To obtain instead $\nu_\mathrm{QPO}\propto \dot M^{3}$, or so, one would have to assume that the size of the precessing torus, and hence its characteristic radius (in units of the truncation radius), $d=R/R_\mathrm{in}$, depends on the mass accretion rate as $\dot M^{-1/2}$, or so, so that 
$$\nu_\mathrm{QPO}\propto d^{-5}\dot M^{3/7}\propto\dot M^{5/2+3/7}\sim\dot M^{3}.$$
It is unknown why the size of the precessing torus should be inversely proportional to the square root, or so, of the mass accretion rate, but as there is no theory of its size this is not impossible to conceive. A stronger objection would be that such a dependence, if universal, would probably destroy the $1/P$ correlation of Eq.~\ref{fit} and Fig.~\ref{FigVibStab}.

On the other hand, as no pulsations have been reported in ULXs displaying the flat-top QPOs, it is possible that they are black holes, and we may try to apply the so-called Lense-Thirring precession model \citep{Fragile+2007,Ingram+Done2011,Teixeira+2014}, in reality involving the $m=-1$ vertical epicyclic mode of a torus \citep{Blaes+2006}. The suggestion of proponents of the model is that the accretion flow changes at some radius from a thin disk to a thick and hot inner torus. This inner torus surrounded by a thinner accretion disk \citep{Bollimpalli24} is supposed to be inclined to the equatorial plane of a Kerr black hole and to precess.
Taking the count rate as a proxy for the X-ray luminosity of a disk truncated (for unknown reasons) at $r=r_0$, we would expect the luminosity to be $L\sim GM\dot M/r_0$. For $\dot M=\mathrm{const}$ this gives $L\propto 1/r_0$, while the ``Lense-Thirring" precession rate of a slender torus is given through first order in the dimensionless Kerr spin parameter $a_*$ by $2\pi\nu=\Omega_\mathrm{K}-\Omega_\perp\approx 2 a_*{(r_g/r)^{3/2}}\Omega_\mathrm{K}\approx 2a_*cr_g^2/r_0^3$ (here, $r_g=GM/c^2$ is the gravitational radius, $\Omega_\perp$ is the vertical epicyclic frequency and $\Omega_\mathrm{K}$ is the test-particle orbital frequency). This would fit the bill, giving $\nu_\mathrm{QPO}\propto 1/r_0^3 \propto L^3$. We stress that the necessary assumption here is that $L\propto 1/r_0$, for X-ray luminosity. This leads to a difficulty. In reality the bolometric luminosity is $\propto \dot M/R_\mathrm{cusp}$, where the position of the inner edge of the precessing accretion torus at $r=R_\mathrm{cusp}$ does not vary strongly with the accretion rate (unlike the transition radius between the outer thin disk and the inner thick torus). For this model to work for flat-top QPOs, the precessing torus would have to radiate mostly outside the 1 to 10 keV range (so that it is the disk that chiefly contributes to the reported count rate), and yet the torus would have to imprint its precession frequency on the X-rays. At the same time, the substantial drop in frequency in the low count rate regime implies that  the precessing torus would have to be large at that time,  $r_0>>R_\mathrm{cusp}$.

Be that as it may, we note that flat-top QPOs detected in ULXs being highly variable in frequency seem to have very different properties from the mHz QPOs detected in PULXs.
\section{Conclusions}

Based on our analysis of the magnetic precession model \citep{Lai99,Pfeiffer+04}, as applied to PULXs, we arrive at the following conclusions.

   \begin{enumerate}
      \item The pulsation period ($P$) and the QPO frequency ($\nu_{\rm QPO}$) in PULXs 
      should be inversely related to each other (for the same fractional size of the precessing part of the disk). We show that the expected QPO frequency is around two orders of magnitude lower than the pulsation frequency, $\nu_{\rm QPO}=\mathrm{a~ few}\times (10^{-3}/P$). The model prediction seems to be indeed supported by the observations of the three known PULX sources with QPOs: M82 X-2, M51 ULX-7, and NGC 7793 P13.
      \item Neither the magnetic dipole of the neutron star, nor its mass accretion rate, can be constrained directly by comparing the mHz QPO frequency with the predictions of the model. Remarkably, neither $\mu$, nor $\dot M$,  appear explicitly in Eq.~\ref{qpo}. 
      {\item While the reported mHz QPO frequencies in the PULXs vary in time by no more than 20\%, and hence may agree with the predicted dependence of Eq.~\ref{qpo} for roughly constant values of the parameters, the QPOs in the flat-top ULXs vary by a factor of at least 2 to 4. Having a very different behavior, the QPOs associated with flat-top noise may have a different origin, and may not even be related to neutron stars.} 
      \item We eagerly await reports of QPO measurements in other PULX sources. If they are also found to be in agreement with our interpretation of the magnetic precession model, it may prove possible to determine the size of the precessing part of the inner disk, and to ultimately constrain the geometry of ULX accretion.
   \end{enumerate}

\begin{acknowledgements}
      This project was funded in part by the Polish NCN grant No. 2019/35/O/ST9/03965. M\v{C} also acknowledges the Czech Science Foundation (GA\v{C}R) grant No.~21-06825X and the support by the International Space Science Institute (ISSI) in Bern, which hosted the International Team project \#495 (Feeding the spinning top) with its inspiring discussions. 
\end{acknowledgements}

\end{document}